\documentclass[letter]{aa} 

\usepackage{graphicx}
\usepackage{txfonts}
\usepackage{tikz,xcolor,hyperref}
\definecolor{lime}{HTML}{A6CE39}
\DeclareRobustCommand{\orcidicon}{%
	\begin{tikzpicture}
	\draw[lime, fill=lime] (0,0) 
	circle [radius=0.16] 
	node[white] {{\fontfamily{qag}\selectfont \tiny ID}};
	\draw[white, fill=white] (-0.0625,0.095) 
	circle [radius=0.007];
	\end{tikzpicture}
	\hspace{-2mm}
}

\foreach \x in {A, ..., Z}{%
	\expandafter\xdef\csname orcid\x\endcsname{\noexpand\href{https://orcid.org/\csname orcidauthor\x\endcsname}{\noexpand\orcidicon}}
}

\begin{document}

   \title{Damping wings in the Lyman-$\alpha$ forest}

   \subtitle{A model-independent measurement of the neutral fraction at $5.4<z<6.1$}

   \author{Benedetta~Spina\inst{1}\fnmsep\thanks{b.spina@thphys.uni-heidelberg.de}\orcidA{},
          Sarah~E.~I.~Bosman\inst{1,2}\orcidB{},
          Frederick~B.~Davies\inst{2}\orcidC{},
          Prakash~Gaikwad\inst{2}\orcidD{}
          \and
          Yongda~Zhu\inst{3,4}\orcidE{}
          }
   \authorrunning{Spina et al.}
   \institute{Institute for Theoretical Physics, Heidelberg University, Philosophenweg 12, D–69120, Heidelberg, Germany
         \and
             Max-Planck-Institut f\"{u}r Astronomie, K\"{o}nigstuhl 17, 69117 Heidelberg, Germany
             \and
                    Steward Observatory, University of Arizona, 933 North Cherry Avenue, Tucson, AZ 85721, USA
                    \and
                        Department of Physics \& Astronomy, University of California, Riverside, CA 92521, USA
             }

   \date{ }
 
  \abstract
    {Recent observations have positioned the endpoint of the Epoch of Reionisation (EoR) at redshift $z \sim 5.3$. However, observations of the Lyman-$\alpha$ forest have not yet been able to discern whether reionisation occurred slowly and late, with substantial neutral hydrogen
    persisting at redshift $\sim 6$, or rapidly and earlier, with the apparent late end driven by the fluctuating UV background.
    Gunn-Peterson (GP) absorption troughs
    are solid indicators that reionisation is not complete until $z=5.3$, but whether they contain significantly neutral gas has not yet been proven.}
    {We aim to answer this question by directly measuring, for the first time, the neutral hydrogen fraction ($x_\mathrm{HI}$) at the end of the EoR ($5 \lesssim z \lesssim 6$) in high-redshift quasars spectra. }
    {For high neutral fractions $x_\mathrm{HI}\gtrsim0.1$, GP troughs exhibit damping wing (DW) absorption extending over $1000$ km s$^{-1}$ beyond the troughs. While conclusively detected in Lyman-$\alpha$ emission lines of quasars at $z\geq7$, DWs are challenging to observe in the general Lyman-$\alpha$ forest due to absorption complexities and small-scale stochastic transmission features.} 
    {We report the first successful identification of the stochastic DW signal adjacent to GP troughs at redshifts $z=5.6$ through careful stacking of the dark gaps in Lyman-$\alpha$ forest (SNR$=6.3$). We use the signal to present a measurement of the corresponding global $x_\mathrm{HI}=0.19\pm0.07$ $(_{-0.16}^{+0.11})$ at $1\sigma$ $(2\sigma)$ at $z=5.6$ and a limit $x_\mathrm{HI}<0.44$ at $z=5.9$.}
    {The detection of this signal demonstrates the existence of substantially neutral islands near the conclusion of the EoR, unequivocally signaling a late-and-slow reionization scenario.}
    
   \keywords{intergalactic medium --
                quasars: absorption lines --
                dark ages, reionization, first stars
               }

   \maketitle
%

\section{Introduction}

The Gunn-Peterson (GP) effect is pivotal to constrain the Epoch of Reionization (EoR), i.e.~the transition of the Intergalactic Medium (IGM) from a predominantly neutral to an ionized state \citep{GP}. The effect is observed as ``GP troughs'' showing nearly complete absorption of rest-frame $1215.67$\AA \ photons by neutral hydrogen with an H{\small{I}} fraction $x_\mathrm{HI} \gtrsim 10^{-4}$ along the lines of sight to background quasars \citep{Fan06,Becker15,Bosman22}. 
GP troughs more than $30$ cMpc in length are found in the spectra of more than half of quasars at $z>5.7$ and persist down to $z=5.3$ \citep{Zhu21}, an observation which rules out a uniformly ionised Universe. However, whether those GP troughs correspond to significantly ``neutral islands'' of gas with $x_\mathrm{HI} \simeq 1$ or to fluctuations in the UV ionising background (UVB) in near-fully ionised gas with $x_\mathrm{HI} \lesssim 0.01$ at $z=6$ is still unknown \citep{Gaikwad23,Davies23}. The most stringent upper limit on $x_\mathrm{HI}$, based on the total absorbed fraction of the Lyman-$\alpha$ (Ly$\alpha$) forest, permits $x_\mathrm{HI}<0.6$ at $z=6$ \citep{McGreer15}, while the most stringent lower limit from the Ly$\alpha$ forest at this redshift is $x_\mathrm{HI} > 7 \times 10^{-5}$ from fluctuations in optical depth \citep{Bosman22}. Models of the end of reionisation between these extremes have until now remained indistinguishable. 

\begin{figure*}
\centering
\includegraphics[width=0.9\hsize]{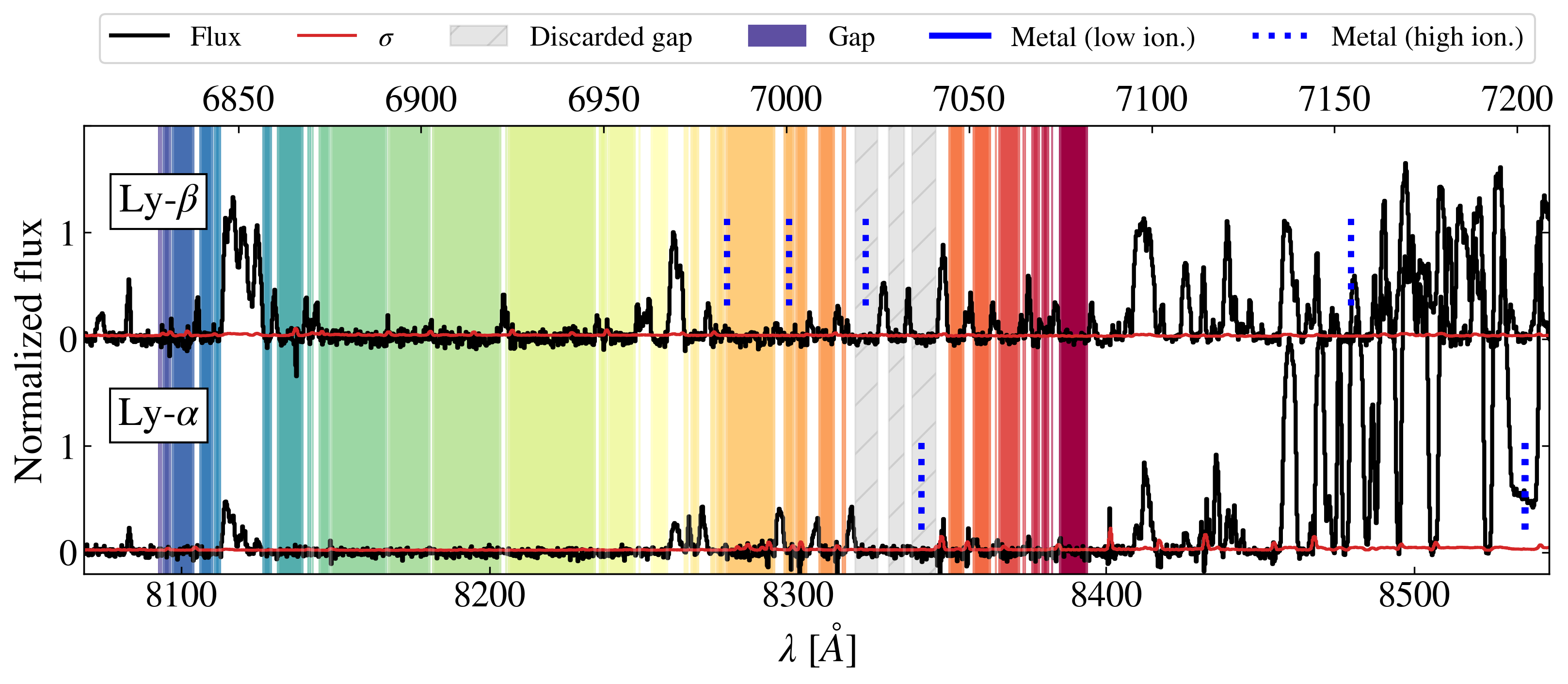}
\caption{Example of the gap-finding procedure for the sight-line towards quasar ATLASJ029.9915-36.5658. The Ly$\alpha$ (bottom) and Ly$\beta$ (top) spectra are depicted along with identified gaps (colored bands). Contamination from metal systems are also displayed (blue vertical lines) and the corresponding gaps removed (gray-strip bands). In this example, a high-ionisation metal system potentially affecting the Ly$\alpha$ forest is identified, leading to the removal of $3$ joint gaps.}
\label{plot1}
\end{figure*}

At high neutral fractions $x_\mathrm{HI}\gtrsim0.1$, GP troughs start exhibiting absorption in damping wings (DW) extending over $1000$ km s$^{-1}$ beyond the troughs themselves. This DW absorption has been detected conclusively over the Ly$\alpha$ emission lines of quasars when the neutral gas is located relatively close to them, giving the first and most robust detections of reionisation-related neutral IGM gas ($x_\mathrm{HI}\sim0.5$) at $z\geq7$ \citep{Mortlock11,Davies18,Greig22}. However, DWs cannot be detected on an individual basis in the general Ly$\alpha$ forest due to (1) significant absorption throughout the forest at $z>5$ and (2) strong small-scale stochastic transmission features (``transmission spikes''). 
In this paper, we present a successful detection of the DW signal around GP troughs at $z=5.6$ and $z=5.9$ from stacking of GP troughs in the Ly$\alpha$ forest, an idea first proposed by \citet{MalloyLidz15}. A detection and measurement of the corresponding necessary global $x_\mathrm{HI}$ is made possible by the great increase in data quality and quantity from the XQR-30 survey \citep{D'Odorico23} and by several improvements to the originally proposed method.

In Section \ref{sec:sample}, we present our observational sample consisting of 38 X-Shooter spectra of quasars at $5.4<z<6.5$. Our method is described in Section \ref{sec:methods}, including GP trough identification (\ref{sec:gap_finding}) and stacking (\ref{sec:gap_stacking}) and $x_\mathrm{HI}$ estimation (\ref{sec:constraints}). We present our measurements of the global $x_\mathrm{HI}$ fraction at $z=5.6$ and a limit at $z=5.9$ in Section \ref{sec:results}. Throughout the paper we use a Planck flat LCDM cosmology \citep{Planck18}, distances are given in comoving units, and $x_\mathrm{HI}$ is volume-averaged.

\section{Sample}
\label{sec:sample}

Our observational sample consists of 38 X-Shooter spectra selected from \citep{Bosman22}. We require the quasars to lie in the redshift range $5.4<z<6.5$ and to have a minimum signal-to-noise ratio  SNR$>20$ per 10 km s$^{-1}$ pixel. The resolution of the spectra is $\sim34$ km s$^{-1}$ over the Ly$\alpha$ forest (see \citealt{D'Odorico23}).
A list of the quasars in the sample with redshifts and SNR is presented in Appendix \ref{sec:QSO}.

The Ly$\alpha$ and Lyman-$\beta$ (Ly$\beta$) forests of the quasars are normalised using reconstructed underlying continua as described in \citet{Bosman22} and \citet{Zhu22}, which employ the same spectral regions of the same spectra. We use a near-linear log-PCA approach based on the method of \citet{Davies18}, with the improvements described in \citet{Bosman21}.  


\begin{figure*}
\centering
\includegraphics[width=0.9\hsize]{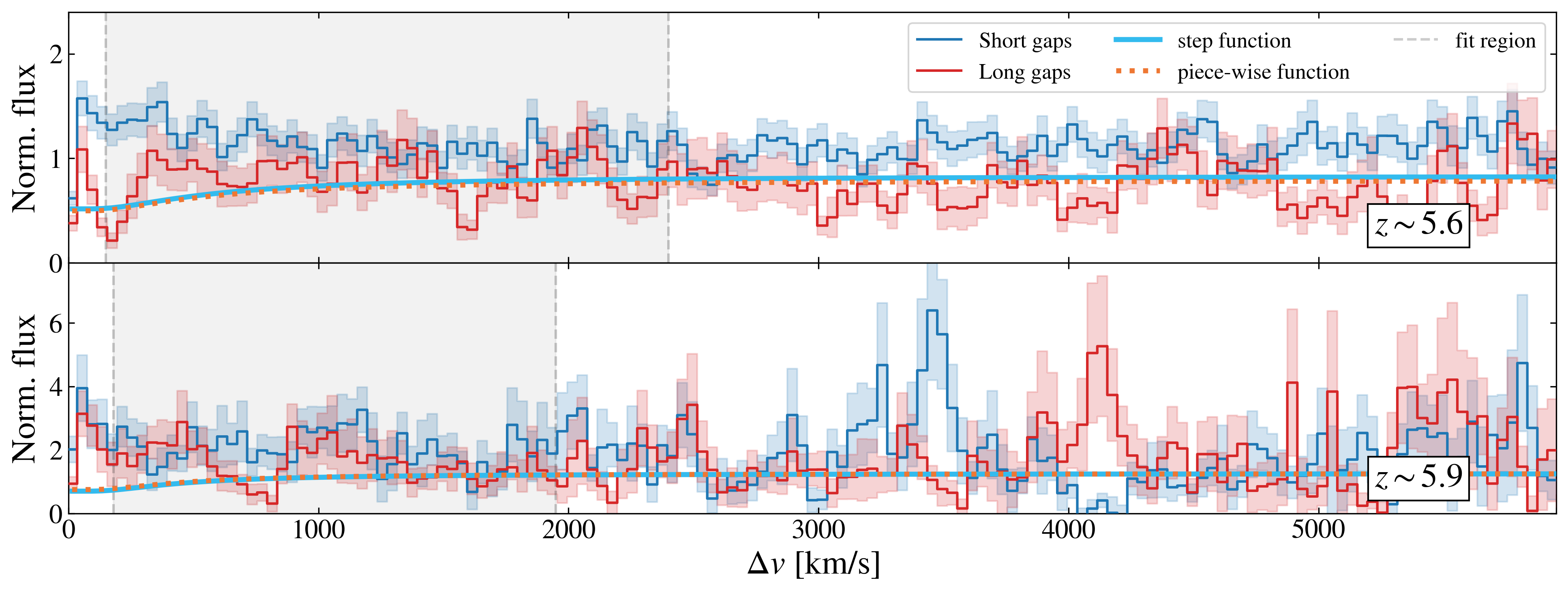}
\caption{Stacked spectrum around long and short gaps for two redshift bins (top: $z=5.6$, bottom: $z=5.9$). In each panel, the red and blue curves depict the stacked profiles of long and short gaps, respectively, with their widths being the uncertainties from bootstrap resampling. The continuous light-blue line and the dotted orange line indicate the best-fit transmission model fit to the stack around long gaps for the step-function model and the piece-wise model, respectively.  Long gaps are defined as those with velocity widths greater than $340$ km s$^{-1}$, while short gaps have lengths under $200$ km s$^{-1}$. The curves in the bottom panel show the models which are permitted at at $2\sigma$ level.}
\label{plot2}
\end{figure*}

\section{Methods}
\label{sec:methods}

The Ly$\alpha$ forest saturates at the mean cosmic density as soon as the neutral fraction of the absorbing gas reaches $x_{\text{HI}} = 10^{-4}$. However, multiple models predict that the GP troughs observed down to $z=5.3$ are highly neutral, with $x_{\text{HI}}>0.1$ \citep{Kulkarni19}. While saturation of the forest on its own is not sensitive in this regime, such large column densities of neutral gas should start to produce damping wings out to large velocity separations, $\Delta v>1000$ km s$^{-1}$ from the GP troughs. While Ly$\alpha$ transmission is too highly stochastic to allow observing DWs around individual troughs, in theory the mean neutral fraction of GP troughs can be recovered through the a detection of the DW in a deep stack of spectra -- a method first suggested theoretically by \citet{MalloyLidz15}. We are now in a position to detect this signal in practice for the first time. We largely follow the method outlined by those authors, with changes specifically noted.

In order to select and stack absorption gaps which potentially contain neutral gas, we search for regions of the forest which are fully absorbed in both Ly$\alpha$ and Ly$\beta$. Those gaps are then stacked using the edge of the Ly$\beta$ trough as a point of alignment. Indeed, transmission in the Ly$\beta$ forest (or any other higher-order forest) indicates that the gas is not opaque at the Lyman limit, i.e.~it cannot constitute a Damped Ly$\alpha$ absorber (DLA). Since accurate methods for reconstructing the intrinsic quasars continua are not available beyond the Ly$\beta$ forest, we do not use higher-order transitions in this paper. 

\subsection{Gap finding}
\label{sec:gap_finding}

First, we exclude a $\sim 5000$ km/s region on the blue side of the Ly$\alpha$ (Ly$\beta$) emission line. This exclusion window is sufficient for all but 3 quasars, where we manually extended the masked region due to unusually long proximity effects. 

In order to identify absorbed regions in both the Ly$\alpha$ and Ly$\beta$ forests, we must contend with (at least) two sources of contamination: sky-line residuals which may artifically ``split'' the gaps, and DLAs which are known to be associated with galactic metal systems rather than the IGM.

Our spectra contain residuals due to the removal of sky-lines. Under the assumption that sky-line residual spikes are narrow, affect only a limited number of adjacent pixels, and the genuine transmission spikes will have an SNR per pixel of at least $2.5$, we clean our signal of sky-line residuals in the following way. First, we smooth the spectrum with a gaussian filter with a width of $2$ pixels, and compare the smoothed and unsmoothed signals. If the difference between the smoothed and unsmoothed signal exceeds 2.5 times the observational uncertainty at any pixel, we consider it to be affected by sky-line residuals. 
We then replace the value of the contaminated pixel with the value of the smoothed signal (re-computed after masking). In this way, pixels most likely to be intrinsically absorbed are assumed to be opaque, and pixels more likely to be transmissive are given flux marking them as not being gaps.

We then identify statistically significant absorbed gaps and determine their edges.
The flux distribution of the cleaned signal reveals two distinct features: firstly, a Gaussian-like distribution centered around zero, representing expected fluctuations in the spectral noise; and secondly, a broader distribution reaching higher flux values, indicative of transmission peaks. 
Utilizing the Gaussian-like noise distribution, we establish a threshold for significant detection, denoted as $\nu$. Peaks in the broader distribution are considered genuine if they exhibit an SNR per pixel of at least $3$ compared to the noise distribution.

Further contamination arises from galactic DLA systems along the line-of-sight, which may be mistaken for Gunn-Peterson troughs. We use the catalog of metal absorbers identified by \citep{DaviesR23} to exclude any gaps which are potentially galactic DLAs. We exclude gaps in Ly$\alpha$ which could be caused by either low-ionisation or high-ionisation systems, and gaps in Ly$\beta$ which could be caused by lower-redshift Ly$\alpha$ absorption from low-ionisation systems. We retain low-redshift, high-ionisation systems which would affect the Ly$\beta$ forest alone, since they are very numerous, typically only have modest Ly$\alpha$ absorption, and their effects do not mimic the signal we are looking for; they only potentially increase the stochastic noise.

Finally, we identify which gaps to stack. We start with all foreground regions which are opaque in both Ly$\alpha$ and Ly$\beta$. To increase the sensitivity of the measurement, we modify the approach of \citet{MalloyLidz15} to use both sides of such gaps (red and blue) since DWs 
are expected to affect both nearly equally.
Further, we try to mitigate the effect of contamination of the Ly$\beta$ forest by overlapping lower-redshift Ly$\alpha$ absorption. We note that there is no physical reason, other than contamination, for the Ly$\beta$ forest to saturate while Ly$\alpha$ does not.  We therefore exclude from the stack all gap edges which are ``terminated'' or ``split'' by Ly$\alpha$ transmission, since that fact makes it clear that the gap has run into highly ionised gas. 

In Figure \ref{plot1}, we illustrate our gap-finding algorithm for a single sightline. The Ly$\alpha$ and Ly$\beta$ spectra are shown with coloured bands indicating identified gaps.
We identify in grey the gaps which were excluded due to corresponding to potential DLAs identified thorugh metal absorption, while the white regions are not gaps (i.e.~transmission spikes, or quasar proximity zone).

\subsection{Gap stacking}
\label{sec:gap_stacking}
We stack the gaps which we identify as outlined above. The gaps are stacked at their edges, ideally corresponding to the H~{\small{I}}/H~{\small{II}} regions limit. Prior to stacking, we normalize the flux using the global mean flux as a function of redshift, using the functional form given in  \citet{Bosman22}. To enhance the robustness of our analysis and expand our sample, we employ a strategy of flipping and stacking the blue edges of the gaps with the red edges.
We compute uncertainties through a bootstrap procedure on the selected gaps. 

We explore the redshift dependence of the H~{\small{I}} fraction by dividing our sample into two redshift bins, $5.4 \leq z < 5.8$ ($\left\langle z \right\rangle = 5.67$) and $5.8 \leq z < 6.1$ ($\left\langle z \right\rangle = 5.89$). Following the methodology established in \citet{MalloyLidz15}, we will begin by comparing stacks of the Ly$\alpha$ forest around ``short'' and ``long'' gaps. The rationale behind this comparison lies in the expectation that short gaps, originating from H~{\small{II}} regions, should contain minimal H~{\small{I}} content. By contrast, we anticipate long gaps to exhibit an excess of H~{\small{I}}. 

Short gaps are defined as those with a velocity width less than $200\,\mathrm{km\,s^{-1}}$, while long gaps is determined by a threshold parameter, $L_\mathrm{thres}$. A higher value of 
$L_\mathrm{thres}$ limits the stack to longer gaps which are more likely to be highly neutral, but at the cost of sample size. In the analysis of the signal below, we will usually marginalise over $L_\mathrm{thres}$ in order to remain agnostic as to the minimum length of gaps which are significantly neutral. 

\begin{figure*}
\centering
\includegraphics[width=0.9\hsize]{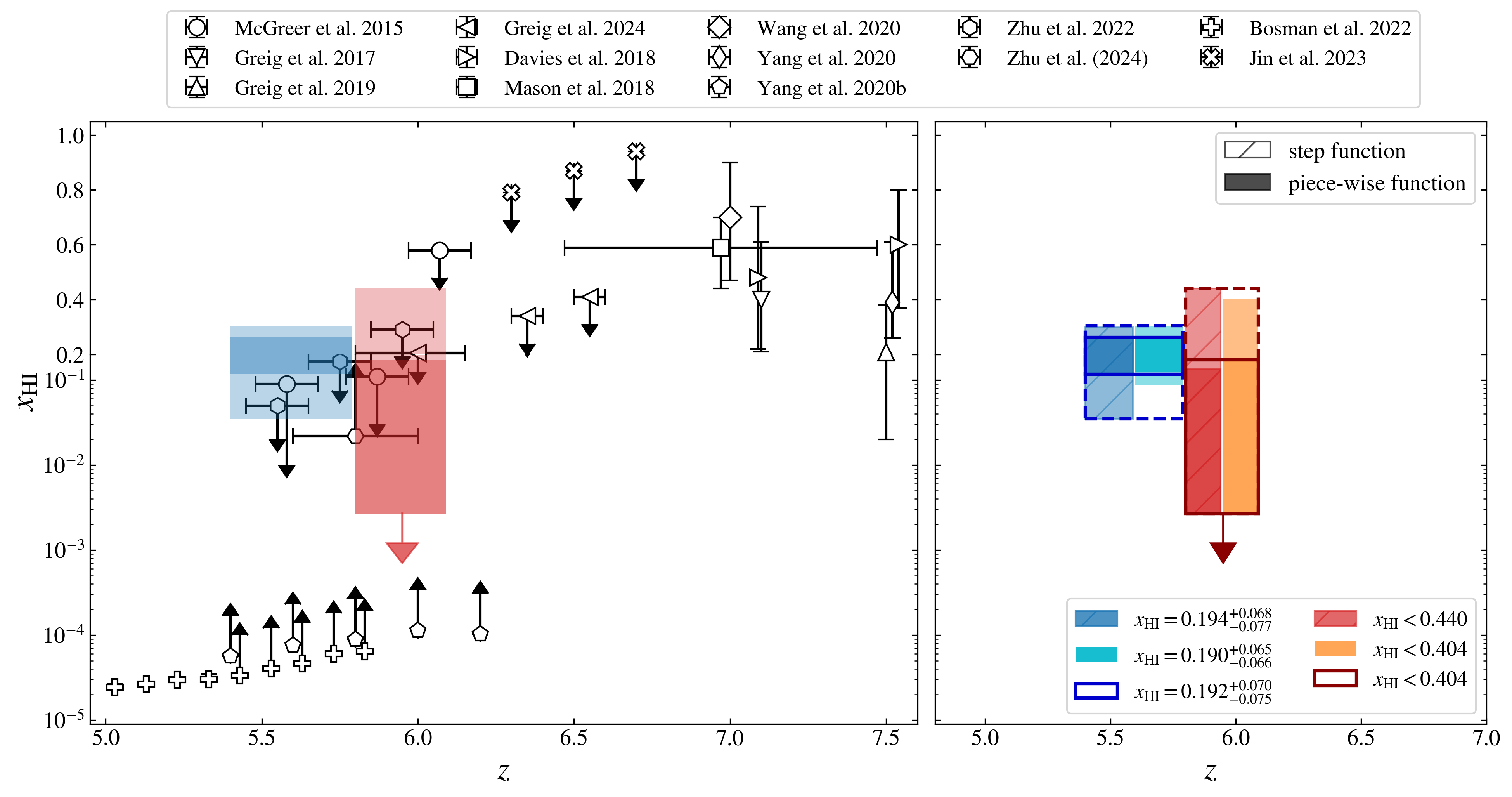}
\caption{Constraints on the neutral fraction across cosmic time. The coloured solid and dashed boxes indicate our $1\sigma$ and $2\sigma$ constraints, respectively. Right: the constraints obtained in this work, in the two redshift bins and with the two $x_\mathrm{HI}$-$L$ relations proposed. Left: existing constraints from the literature: \citet{McGreer15}, \citet{Greig17}, \citet{Greig19}, \citet{Davies18}, \citet{Mason18}, \citet{Mason19},   \citet{Wang20}, \citet{Yang20}, \citet{Yang20b},  \citet{Zhu22}, \citet{Bosman22}, \citet{Jin23}, \citet{Zhu24}. Some of the literature points have been slightly shifted in redshift, and the constraints of \citet{Greig24} have been combined for improved clarity.} 
\label{plot3}
\end{figure*}

As will be shown in Section \ref{sec:results} and Figure \ref{plot2}, we achieve a detection of the statistical DW signal in the lower-redshift bin at $z=5.6$. To interpret this signal without relying on existing model-dependent simulations, we develop a simple analytical framework to estimate the most likely value of the global $x_{\text{HI}}$ giving rise to the observations.

\subsection{Constraints on the H{\small{I}} fraction}
\label{sec:constraints}

We employ two toy models to capture the distribution of neutral gas among gaps of varying lengths. 
The simplest model is a step-function relation, i.e., gaps longer than a threshold $L_C$ originate from completely neutral hydrogen, while shorter gaps arise from H~{\small{II}} regions. The local H~{\small{I}} fraction $\mathrm{x}_\mathrm{HI}^L$ of such gaps is given by
\begin{equation}
    \mathrm{x}_\mathrm{HI}^L \left(L,L_C\right) = 
    \begin{cases}
        0, & L<L_C, \\
        1 & L\geq L_C.
    \end{cases}
\end{equation}

In an attempt to make the distribution more realistic, we also explore a piece-wise relation.  In this model, we allow for a mixture of neutral and ionized hydrogen clouds to populate gaps. The H~{\small{I}} fraction, $\mathrm{x}_\mathrm{HI}^L$, of gaps is then given by
\begin{equation}
    \mathrm{x}_\mathrm{HI}^L \left(L,L_C\right) = 
    \begin{cases}
        0, & L<200\,\mathrm{km/s}, \\
        \dfrac{L-200}{L_C-200} & 200\,\mathrm{km/s} \leq L < L_C, \\
        1 & L\geq L_C.
    \end{cases}
\end{equation}
We based this model on the findings of \citet{Zhu22}, where a similar functional form was the best explanation for the observed trough lengths in the Ly$\alpha$ and Ly$\beta$ forests.

We allow the mean flux $T$ to vary. Since the gaps are normalized by the mean flux at the redshift considered, such a free parameter should not be necessary. However, the strong correlation of the Ly$\alpha$ forest on very large scales around gaps of different lengths prevents the mean flux from converging to 1, as we will discuss in \ref{sec:results}. The parameter $T$ is dominated by the behaviour of the stacked flux at large velocity separations and is tightly constrained, so we hold it fixed at its optimal value rather than marginalising over it in the analysis.

Both of these simple models contain only 1 parameter, $L_C$, which is the gap length above which full neutrality of the gas is assumed. For each value of $L_C$, we generate the shape of the expected stacked DW via
\begin{equation}
    \left\langle F \right\rangle = \left[ \left( 1 - \mathrm{x}_\mathrm{HI}^L \right) e^{-\tau_D \left( \mathrm{x}_\mathrm{HI}^L = 0 \right)} + \mathrm{x}_\mathrm{HI}^L e^{-\tau_D \left( \mathrm{x}_\mathrm{HI}^L = 1 \right)} \right] T
\end{equation}
where $\tau_D(x_{\text{HI}})$ is the conventional Gunn-Peterson damping wing optical depth \citep{Mortlock16}. 

We fit this functional shape to the signal using $\chi^2$ minimisation. We make use of the stacked signal over a velocity range $150 < \Delta v < 3000$ km s$^{-1}$ at $z=5.6$ and $180 < \Delta v < 3000$ at $z=5.9$, but we do not use the signal beyond the $\Delta v$ where less than $100$ spectra are stacked. 
Since the covariance matrix of the stacked signal contains strong non-diagonal elements which prevent inversion, we perform parameter inference empirically instead. We perform 5000 draws of the stacked signal via bootstrap and obtain posteriors on $L_C$ from the distribution of their optima. We then repeat this procedure for a variant of the signal using a different $L_\text{thresh}$, the lower limit for a ``long'' gap. Finally, we marginalise over $L_\text{thresh}$ to obtain the final posterior on $L_C$.

The central value of $L_C$ and its uncertainties can be translated into a global neutral fraction $x_{\text{HI}}$. To do this, we assign to each gap used in the analysis its own local neutral fraction according to Eqs.~(1) or (2), and we assign a value $x_{\text{HI}}=0$ to the un-masked, non-gap portion of the quasar sightlines. The resulting volume-average is our estimate of of the global $x_{\text{HI}}$.

\section{Results}
\label{sec:results}

\subsection{Detection of a statistical damping wing signal}

By comparing the stacked Ly$\alpha$ flux around short and long gaps at $z=5.6$, we detect a very strong signal at $\Delta v < 500$ km s$^{-1}$ which is consistent with qualitative expectations from a statistical DW (Fig.~\ref{plot2}). The decrement is significant at SNR$=6.3$, as determined by re-binning the signal on a matching scale of $80$ km/s and computing the bootstrap uncertainties on the range. The SNR does not scale well with binning, since the source of the uncertainties is predominantly in cosmic variance. We repeat this procedure for the rest of the stack over the usable velocity range, and find no other features significant at SNR$>4$. We show the SNR distributions in Appendix B.

The up-turn in transmission around H~{\small{II}} regions, predicted by \citet{MalloyLidz15}, occurs because the edges of Ly$\alpha$ and Ly$\beta$ gaps align when the IGM is ionized and over-dense. This results in the first pixel of the Ly$\alpha$ stack being above the mean flux. The detection of this feature in small gaps confirms they contain mostly ionized gas, while its partial presence in long gaps suggests these gaps either contain ionized gas or are caused by foreground contamination from lower-redshift Ly$\alpha$ absorption in the Ly$\beta$ forest. 

However, other features and differences are also visible in stacked flux, which were not predicted by e.g.~the forecasts of \citet{MalloyLidz15}. In the central $\sim3$ velocity pixels, corresponding to $\Delta v<150$ km s$^{-1}$, the flux stacks around both the long and short gaps show a clear excess compared to larger separations. We suspect that this is due a large fraction of the gaps having a very low $x_{\text{HI}}$, causing the edges of the Ly$\alpha$ and Ly$\beta$ gaps to overlap. For such gaps, the next pixels past the edge of the gap are far more likely to be a spike than in the general IGM, causing the stacked signal to shoot up above the mean. The fact that this boost persists to $\sim500$ km s$^{-1}$ in the stack around short gaps indicates that transmission spikes are likely clustered, something which has been theorised and predicted to occur on exactly these scales \citep{Wolfson23}.

Another unexpected effect is that the mean fluxes around short gap and around long gaps do not converge even up to $\Delta v=6000$ km s$^{-1}$, corresponding to a distance of 62 Mpc. This is indicative of extremely long correlation scales in Ly$\alpha$ transmission at the end of the EoR, as hinted at by the existence of individual GP troughs over $110$ Mpc in length \citep{Becker15}. Long gaps are more likely to be located in sightlines which are more opaque than average on large scales.

No difference between the flux stack around short gaps and long gaps is visible at $z=5.9$. The mean transmission at this redshift is $4$ times lower than at $z=5.6$, as only $\sim1\%$ of quasar emission is transmitted in the forest \citep{Bosman22}. Despite our analysis containing more gaps and gap edges at $z=5.9$, the SNR of the stack is most likely not sufficient to detect the signal. We use the $z=5.9$ stack to pose a limit on $x_\text{HI}$.

A similar signal was seen independently by \citet{Zhu24}.

\subsection{Implications for the neutral fraction}

We use our analytical framework as a first estimate of the rough global $x_{\text{HI}}$ required by the strength of the signal. To achieve this, we mask the central $150$ km s$^{-1}$ of the signal and allow the mean transmission at large separations to vary ($T$), as outlined in the previous section. The best fits to the $z=5.6$ signal is shown in the top panel of Figure \ref{plot2} as a solid blue and dashed orange curves, corresponding to a step function and a piecewise function distribution neutral gas as a function of gap length. While the two models agree well, they are both relatively poor fits to the shape of the signal. Nevertheless, we propagate the uncertainties to achieve a measurement of $x_\text{HI} = 0.19 \pm 0.07$ $(_{-0.16}^{+0.11})$ at $1\sigma$ $(2\sigma)$ by taking the mean and the outer envelope of constraints from the step and piecewise functions. We briefly investigate the effect of the maximum fraction of gaps which are neutral being dropped to 75\% in the piece-wise assumption: we then would recover a value of $x_\text{HI}$ which is $\sim15\%$ higher (but comfortably within our uncertainties). At $z=5.9$, the posterior returns an upper limit on $x_{\text{HI}} < 0.44$ at $2\sigma$.

We compare these values to the literature in Figure \ref{plot3}. While the $z=5.9$ limit is broadly consistent with other tracers, the value we find of $x_\text{HI}\sim0.2$ at $z=5.6$ is in mild tension with evidence from dark pixels \citep{McGreer15} and the distribution of dark gap lengths \citep{Zhu22}. A neutral fraction of $20\%$ at $z=5.6$ is higher than predicted by models even with ultra-late reionisation \citep{Kulkarni19,Asthana24}. We note that, although the shape of the signal appears poorly captured by our simple analytic form, the strength of the signal is stronger than our fit (Fig.~\ref{plot2}), especially when compared to the stack around short absorption gaps.

However, the extra features observed in the $z=5.6$ spectral stack suggest that our simple analytical model might not be sufficient to characterise the signal and extract a global neutral fraction. In future work, we will employ simulation suites to hopefully model the signal and its contamination fully. Still, we note that even the largest existing cosmological simulations struggle to reproduce the very large scales over which coherence in the mean flux is seen, while maintaining sufficient resolution to reproduce the clustering of transmission spikes. We are looking forward to more detailed modelling in future work.

\section{Summary}

We conducted a search for damping wings inside the Ly$\alpha$ forest at the end of the EoR. Following the method laid out by \citet{MalloyLidz15}, we identified  IGM regions opaque to both Ly$\alpha$ and Ly$\beta$ which may therefore contain significantly neutral gas. 

We stacked the Ly$\alpha$ forest past the edge of these dark gaps, as identified by the end of the trough in Ly$\beta$. Following \citet{MalloyLidz15}, we compare the stacked flux seen around short gaps (length $<200$ km s$^{-1}$) and longer gaps (with a variable threshold; $340$ km s$^{-1}$ is shown in Figure \ref{plot2}). We find a clear signal consistent with a stacked DW on scales $\sim 300$ km s$^{-1}$ around the long gaps and not the short gaps at $z=5.6$, but no such signal at $z=5.9$ where our SNR is significantly lower.

In addition to the DW signal, the stacked flux around the long gaps also displays two unexpected features: a boost in the very central pixels, which we attribute to contamination by highly ionised gaps, and a systematic shift towards a lower mean flux compared to the short gaps on very large scales $>60$ Mpc, probably due to large-scale correlations in the UVB. These features were not predicted by forecasts of this signal.

We nevertheless attempt to model the stack around long gaps after masking the central contamination boost. We assign to each gap a neutral fraction as a function of its length via two toy models which give consistent results. The resulting best fit implies a high neutral fraction of $x_\text{HI} = 0.19 \pm0.07$, or $x_\text{HI} > 0.03$ at $2\sigma$. At $z=5.9$, we can only pose an upper limit $x_\text{HI} < 0.44$.

While more modeling is needed to understand the complexities of the stacked damping wing signal, the fact that it is detected at all in the $z=5.6$ IGM is an unequivocal sign of the existence of significantly neutral islands at the end of the EoR. 

\begin{acknowledgements}
The authors thank the helpful anonymous referee who contributed valuable comments which improved the manuscript.
BS and SEIB are supported by the Deutsche Forschungsgemeinschaft (DFG) under Emmy Noether grant number BO 5771/1-1.
Part of this work is based on observations collected at the European
Southern Observatory under ESO programme 1103.A-0817. This work made use of the publicly available software package \textsc{CoReCon} \citep{Garaldi23}.
\end{acknowledgements}

\bibliographystyle{aa}
\bibliography{aanda}

\begin{appendix}

\section{QSO list}
\label{sec:QSO}

We report in Table the sample selected for the analysis, selected from \citet{Bosman22}.

\begin{table}[h]
\centering
\begin{tabular}{l l l c}
Quasar ID & $z_{\text{qso}}$ & $\mathrm{SNR\,pix^{-1}}$ \\
\hline
\hline
J0108+0711          & $ 5.577$   & $20.0$   \\
J1335-0328          & $ 5.693$   & $35.0$   \\
PSO J215-16         & $ 5.7321$  & $30.2$   \\
SDSS J0927+2001     & $ 5.7722$  & $53.8$   \\
PSO J308-27         & $ 5.7985 $ & $53.2$   \\ 
SDSS J0836+0054     & $ 5.804$   & $73.8$   \\
PSO J065+01         & $ 5.833 $  & $25.1$   \\
PSO J242-12         & $ 5.837 $  & $22.9$   \\
PSO J025-11         & $ 5.844 $  & $50.6$   \\
PSO J183-12         & $ 5.917 $  & $61.8$   \\
PSO J108+08         & $ 5.9485 $ & $104.8$  \\
PSO J056-16         & $ 5.9676$  & $32.0$   \\
PSO J029-29         & $ 5.984 $  & $65.6$   \\
SDSS J0818+1722     & $ 5.997$   & $132.1$  \\
ULAS J0148+0600     & $ 5.998$   & $152.0$  \\
PSO J340-18         & $ 5.999$   & $29.9$   \\
PSO J007+04         & $ 6.0015 $ & $54.4$   \\
SDSS J2310+1855     & $ 6.0031 $ & $113.4$  \\
ATLAS J029-36       & $ 6.021 $  & $57.1$   \\
SDSS J1306+0356     & $ 6.0330$  & $65.3$   \\
VDES J0408-5632     & $ 6.0345 $ & $86.6$   \\
ULAS J1207+0630     & $ 6.0366$  & $29.2$   \\
ATLAS J158-14       & $ 6.0685 $ & $60.3$   \\
SDSS J0842+1218     & $ 6.0754 $ & $83.2$   \\
PSO J239-07         & $ 6.1102 $ & $56.3$   \\
CFHQS J1509-1749    & $ 6.1225$  & $43.0$   \\
ULAS J1319+0950     & $ 6.1347$  & $81.7$   \\
PSO J217-16         & $ 6.1498 $ & $73.0$   \\
PSO J217-07         & $ 6.1663 $ & $33.3$   \\
PSO J359-06         & $ 6.1722 $ & $68.8$   \\
PSO J065-26         & $ 6.1871 $ & $77.9$   \\
PSO J060+24         & $ 6.192 $  & $49.7$   \\
PSO J308-21         & $ 6.2355$  & $24.4$   \\
SDSS J1030+0524     & $ 6.309$   & $69.6$   \\
VST-ATLAS J025-33   & $ 6.318$   & $127.3$  \\
SDSS J0100+2802     & $ 6.3269$  & $560.5$  \\
ATLAS J2211-3206    & $ 6.3394 $ & $37.5$   \\
PSO J159-02         & $ 6.386$   & $22.9$   \\
DELS J1535+1943     & $ 6.3932 $ & $22.6$   \\
PSO J1212+0505      & $ 6.4386 $ & $55.8$   \\
\hline
\end{tabular}
\caption{Sample from \cite{Bosman22}.} 
\label{table:XQR30}
\end{table}

   \begin{figure*}
   \centering
   \includegraphics[width=\hsize]{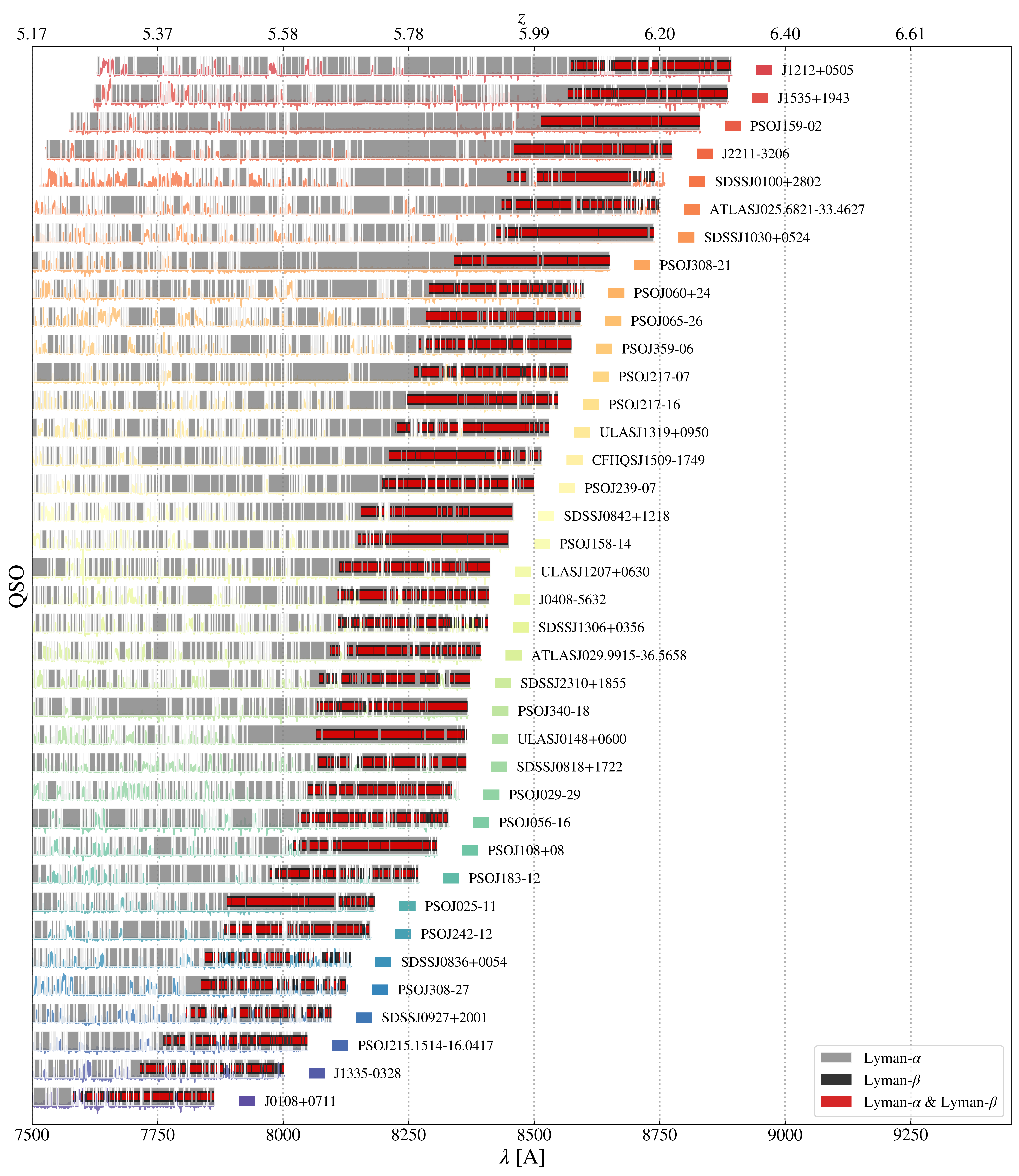}
      \caption{Overview of dark gaps identified in the Ly$\alpha$ and Ly$\beta$ forests from our sample of 38 QSO spectra. Light (dark) grey shaded regions identify the gaps in the Ly$\alpha$ (Ly$\beta$) region. The red shaded regions show the combination of such gaps.}
         \label{QSO}
   \end{figure*}

\section{Robustness of the DW signal}

\begin{figure*}
\centering
\includegraphics[width=\hsize]{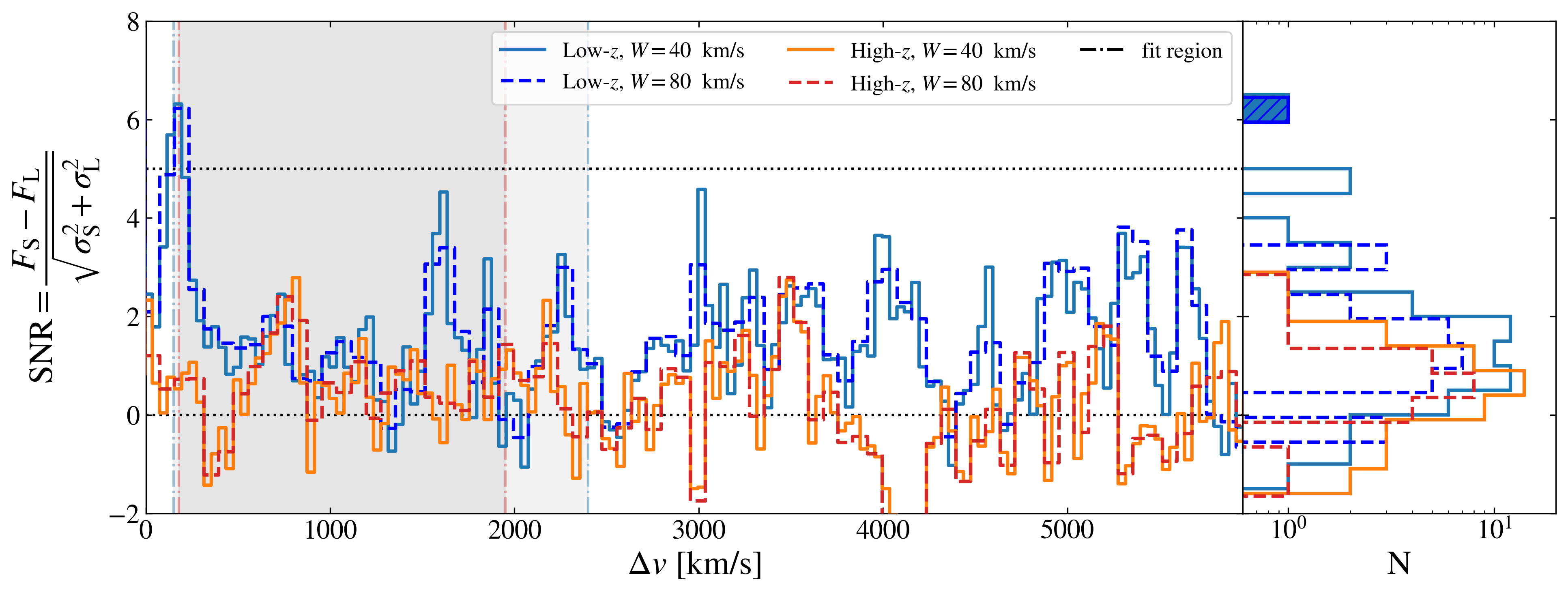}
\caption{Significance of the flux decrement near $\Delta \nu = 0$. In the left panel, the signal-to-noise ratio per pixel is illustrated for different binning windows for the two redshift bins. The gray shadow area corresponds to the fit region, while $F_S$ ($\sigma_S$) and $F_L$ ($\sigma_L$) represent the (errors on the) flux of the short and long gaps respectively. The right panel shows the significance distribution within the (gray) fit region. The pixels with a SNR larger than 5 are highlighted.}
\label{plot4}
\end{figure*}

We test the robustness of the DW signal by looking at the difference between the flux in the long and short gap stacks close to $\Delta \nu = 0$. In Figure \ref{plot4}, we display the significance of such flux decrement with different binning windows ($W = 40\, \mathrm{km\, s^{-1}}$, as for our nominal bining scale, and $W = 80\,\mathrm{km\, s^{-1}}$) for the two redshift bins. Our analysis shows that at $z\sim 5.6$ there is not another large-flux decrement in the spectrum similar to the one near $\Delta \nu = 0$. A signal-to-noise ratio per pixel of $\gtrsim 6$ is found only for $180\, \mathrm{km\,s^{-1}} \lesssim \Delta \nu \lesssim 260\, \mathrm{km\,s^{-1}}$ at $z\sim 5.6$. 

\end{appendix}

\end{document}